\documentclass[12pt, preprint]{aastex}

\slugcomment{Submitted to the Astronomical Journal, August 7, 2007}
\shorttitle{Candidates and Age of KBO 2003 EL61 Family}
\shortauthors{Ragozzine \& Brown}

\title{Candidate Members and Age Estimate of the Family of Kuiper Belt Object 2003 EL61}
\author{D. Ragozzine and M.~E. Brown}
\affil{Division of Geological and Planetary Sciences, California Institute
of Technology, Pasadena, CA 91125}

\email{darin@gps.caltech.edu}

\begin{document}

\newcommand{\elso}{2003~EL61 }
\newcommand{\elsons}{2003~EL61}
\newcommand{\Nfive}{2005 CB79 }
\newcommand{\Kfive}{2005 UQ513 }
\newcommand{\x}{\times}
\newcommand{\e}[1]{10^{#1}}
\newcommand{\be}{\begin{equation}}
\newcommand{\ee}{\end{equation}}
\newcommand{\mpers}{m s$^{-1}$}
\newcommand{\mpersws}{m s$^{-1}$ }

\newcommand{\bfam}{B07 }
\newcommand{\bfamnat}{2007Nature..446..296}
\newcommand{\bfamns}{B07}

\newcommand{\fyninet}{1}
\newcommand{\barkumet}{2}
\newcommand{\bfamt}{3}
\newcommand{\mbosst}{4}
\newcommand{\nollpostert}{5}
\newcommand{\teglert}{6}


\begin{abstract}

The collisional family of Kuiper belt object (KBO) \elso opens the possibility for many interesting new studies of 
processes important in the formation and evolution of the outer solar system. As the first family in the Kuiper belt, 
it can be studied using techniques developed for studying asteroid families, although some modifications are necessary. 
Applying these modified techniques allows for a dynamical study of the \elso family. The velocity required to change 
orbits is used to quantitatively identify objects near the collision. A method for identifying family members that have 
potentially diffused in resonances (like \elsons) is also developed. Known family members are among the very closest 
KBOs to the collision and two new likely family members are identified: 2003 UZ117 and 1999 OY3. We also give tables of 
candidate family members which require future observations to confirm membership. We estimate that a minimum of $\sim$1 
GYr is needed for resonance diffusion to produce the current position of \elsons, implying that the family is likely 
primordial. Future refinement of the age estimate is possible once (many) more resonant objects are identified. 
The ancient nature of the collision contrasts with the seemingly fresh surfaces of known family members, suggesting that our understanding of outer solar system surfaces is incomplete.

\end{abstract}

\section{Introduction} 

The collisional history of Kuiper belt objects (KBO) strongly constrains the formation and evolution of the Kuiper 
belt. The recent discovery by Brown et al. (2007b, hereafter \bfamns), of a KBO family created by a collision promises new 
and valuable information about the early outer solar system. As the first bona fide collisional family 
in the Kuiper belt, it merits further study and comparison to the Hirayama collisional families in the asteroid belt. 
The goal of this work is to identify potential fragments for future observations and study by properly adapting 
techniques used in dynamical studies of asteroid families. The distribution of family members is then be used to estimate the age of the family.

Members of the \elso family are identified by infrared spectra with strong water ice absorptions (\bfamns), as seen in 
large KBO (136108) \elso and its brightest moon \citep{2006ApJ...640L..87B,2007ApJ...655.1172T}. Based on light curve 
observations, \elso appears to be rapidly rotating with an inferred density of at least 2.6 g cm$^{-3}$ 
\citep{2006ApJ...639.1238R}, suggesting that a giant impact stripped roughly 20\% of the icy mantle from the 
proto-\elsons, based on an assumed initial density of about 2 g cm$^{-3}$ as measured for other large KBOs (e.g. Pluto, 
Triton). The identification of the \elso collisional family by \bfam is presumably the discovery of icy mantle 
fragments ejected from this giant impact. The probability of such a giant impact in the current Kuiper belt is 
extremely low, implying that the collision probably occurred in the early history of the outer solar system before any 
significant depletion in mass, as we discuss below \citep{2007astro.ph..3558M}.

\section{Dynamics}\label{dyn}

\subsection{Differences between KBO and Asteroid Families}

The spread in orbital elements created by a collision in the Kuiper belt is different than the previously studied asteroid belt. In both cases, a collision powerful enough to create a family launches fragments at velocities high enough to be gravitationally ejected. The ejection velocity, $\Delta v$, is the velocity at infinity of unbound fragments and typically scales with 
the escape velocity from the surface of the target. Since known KBOs (especially \elsons) are much larger than typical 
asteroids, the expected $\Delta v$ of a family-forming collision is much larger in the Kuiper belt. In addition, the 
typical orbital velocities ($v_{orb}$) in the Kuiper belt are $\sim$5 km s$^{-1}$, compared to $\sim$20 km s$^{-1}$ 
typical of the asteroid belt. The ratio of $\Delta v / v_{orb}$ roughly determines the size of the spread in orbital 
elements that will be achieved by collisional dispersion. Hence, asteroid families will (initially) be far more tightly 
clustered in proper orbital element space than Kuiper belt families. Figure 1 shows the cloud of orbital elements 
created from a velocity dispersion of only 150 \mpersws from a potential origin of the \elsons-forming collision (described 
in more detail below). 

Asteroid families are often identified by looking only at clusters in proper orbital elements. This fails to identify 
true families in the Kuiper belt since the large anticipated spread in orbital elements is typically larger than the 
natural separation between objects \citep{2003EM&P...92...49C}. The \elso family was only identified because family 
members shared a unique spectral signature, in addition to being dynamically clustered. (\bfamns)

Due to interactions with resonances, the dynamical clustering of collisional families grows weaker with time. In the 
asteroid belt, proper element dispersion is aided by drifting semi-major axes due to the Yarkovsky effect acting on 
small bodies \citep{1999Sci...283.1507F}. In the Kuiper belt, objects are generally very stable over the age of the solar system except near mean-motion resonances with Neptune and/or when perihelia drop below about 35 AU
\citep{2002AJ....124.1221K}.

\subsection{Determining $\Delta v$}

To identify dynamically-nearby KBOs for further investigation, we use $\Delta v$ (the required ejection velocity) as a quantitative measurement of 
dynamical proximity. After a collision, each (unbound) fragment assumes a different heliocentric orbit, all of which intersect at the location of the 
collision $(x_c,y_c,z_c)$. At this location, each family member has a different velocity, $\mathbf{v}_{c}$. By definition, $\Delta v$ is the length of 
$(\mathbf{v}_{c} - \mathbf{v}_{c,cm})$, where $\mathbf{v}_{c,cm}$ is the velocity at the collision location of the center of mass orbit, i.e. all 
values of $\Delta 
v$ are measured with respect to the center of mass orbit. After the collision the center of mass orbit is nearly the orbit of the largest fragment; we 
approximate the center of mass orbit with the post-impact orbit of \elsons.  Finding $\Delta v$ would be trivial if we knew the full set of orbital 
elements ($a,e,i,\Omega,\omega,M$) for each object and \elso at the time of the collision. Unfortunately, after a relatively short time, the coherence 
of the original orbital angles is lost and at the present epoch only the proper semi-major axes, proper eccentricities, and proper inclinations are 
known.

Even so, it is possible to use the distribution of proper elements of family members to estimate the orbital angles of 
the center of mass orbit (see below). Once these orbital elements are fully proscribed, there is enough information 
to calculate $\Delta v$ using the additional constraint that all orbits pass through the collision location. For 
asteroid families, Gauss' equations are then used to compute the components of $\Delta v$ (e.g. 
\citet{1995Icar..118..132M}). These equations are only accurate if $\Delta v \ll v$ for each component of $v$, whereas 
for the \elso family, which has a large velocity dispersion ($\gtrsim 200$ \mpers) and a relatively small orbital velocity ($\sim 
4500$ \mpers), Gauss' equations can lead to inaccuracies which can be avoided through a more direct calculation.

Instead, we convert the center of mass orbital elements to Cartesian coordinates, giving the collision location 
$(x_c,y_c,z_c)$ and velocity $\mathbf{v}_{c,cm}$. We then use a direct mathematical conversion of the 
KBO proper elements ($a_P,e_P,i_P$) and the collision location ($x_c,y_c,z_c$) to find the orbital velocity $\mathbf{v}_c$. 
However, these six variables do not uniquely determine the velocity; there are four possible solutions to this 
inversion. (This degeneracy results from two different occurrences where either the positive or negative square root 
can be used.) That is, a single location ($x_c,y_c,z_c$) can be identified by identical orbital elements $a_P$, $e_P$, 
and $i_P$ and four sets of velocities, as there is no way to distinguish between reflections along the line of apsides 
or the line of nodes. There is no \emph{a priori} way to resolve this degeneracy since the information about the original 
orbital angles $(\Omega,\omega,M)$ is lost. Since the goal of this study is to identify all KBOs that could 
potentially be members of the \elso family, we take the smallest value of $\Delta v$. In this way, all KBOs with proper orbital elements that could be dynamically near the collision are identified.

To determine $\Delta v$ of candidate family members requires the orbital elements of the center of mass orbit. This orbit is usually taken from the proper orbital elements of the largest 
fragment with orbital angles chosen to match the distribution of family members in $a_P$-$e_P$-$i_P$ space (see Figures 1 and 2). The longitude of the ascending node ($\Omega_{cm}$) has no effect on this distribution and is ignored. We use the orientation of the collisional cloud in $a_P$ vs. $e_P$ to find the mean anomaly ($M_{cm}$); the argument of perihelion ($\omega_{cm}$) has no effect here. However, $\omega_{cm}$ does change the extent of the inclinations attainable with a particular value of $\Delta v$. In particular, for collisions that occur on the ecliptic ($\omega_{cm} + M_{cm} \simeq 0^{\circ}$ or $180^{\circ}$), require the lowest $\Delta v$ to change the inclination. At the highest and lowest points of the orbit it is very difficult for a collision to change the inclination (apparent from the form of Gauss' equation for inclination changes, see \citet{1995Icar..118..132M}). There is not enough information in the distribution of proper elements of the family members to uniquely determine the component of $\Delta v$ out of the plane of the orbit. In order to proceed, we choose to work with the minimum possible $\Delta v$, which we will call $\Delta v_{min}$. (The analysis in \bfam did not appreciate this aspect of collisional orbit changes and assumed that $\Delta v_{min}$ was the actual escape velocity of the fragments, which may not be true.) The component of $\Delta v$ out of the plane of the orbit is larger by a factor of $\sim2$ on average (i.e. with a randomly chosen $\omega_{cm}$) and at extreme points in the orbit the acutal ejection velocity could theoretically be $\sim5$ or more times greater than $\Delta v_{min}$. However, collisions are most probable near the ecliptic (where the number density is highest), so the correction is probably much smaller. Furthermore, if an isotropic ejection of fragments is assumed, then this ambiguity is removed and the $a_P$-$e_P$-$i_P$ distribution is sufficient to determine the typical ejection velocity \citep{2006Icar..183..296N}. The only way the actual ejection velocities of all the family members could have been significantly greater than $\sim150$ \mpersws is if they all left in a collimated jet in a particular direction from a non-ecliptic collision. Finally, we note that a similar correction factor will apply to most KBOs, roughly preserving the overall ranking of KBOs by dynamical proximity to the collision. 

\section{Potential Family Members}\label{mems}
\subsection{Collision Center}

For the \elso family, we determined the center of mass orbit based on the orbit of \elsons, the largest fragment. However, \elso has diffused from its 
original location due to interaction in the 12:7 mean-motion resonance with Neptune (\bfamns) and its proper elements have changed. Over long timescales, overlapping sub-resonances can cause diffusion of proper eccentricity and inclination \citep{2001Icar..150..104N, 
1997AJ....114.1246M}. For KBOs starting with orbital elements near the center of the \elso family, we have found 
empirically that the chaotic diffusion nearly conserves the proper Tisserand parameter with respect to Neptune, the 50 
MYr average of the osculating Tisserand parameter:
\be
T = \frac{a_N}{a} + 2 \cos(i-i_N) \sqrt{ (a/a_N)(1-e^2) } 
\ee 
where $a_N$ and $i_N$ are the osculating semi-major axis and inclination of Neptune. In particular, 
we can estimate possible past locations of \elso by changing the eccentricity and inclination to preserve the proper 
Tisserand parameter of \elso ($a_p$=43.10 AU and $T_p$=2.83).

The pre-diffusion orbit of \elso is estimated by minimizing the sum of 
$\Delta v_{min}$ for previously identified family members (1995 SM55, 1996 
TO66, 2002 TX300, 2003 OP32, and 2005 RR43) while fixing $a_P=a_{P,EL61}$ and $T_P=T_{P,EL61}$ and allowing the other orbital elements to vary. This 
results in a nominal collision location at $(a,e,i$,$\omega,M)_{cm}$ = (43.10 AU, 0.118, 28.2$^{\circ}$,
 270.8$^{\circ}$, 75.7$^{\circ}$) which is used to generate Figure 1 and the values of $\Delta v_{min}$ for family members 
listed in Table 1. (As expected, the minimal velocities are attained for values of $\omega_{cm}$ that place the collision near the ecliptic.) The results that follow are not strongly dependent on this particular choice of the center of mass orbit. Exploring 
other center of mass orbits (such as the average of proper elements of non-resonant family members) indicates that the exact 
values of the ejection velocity vary somewhat, especially for objects near ($\Delta v \lesssim 100$ \mpers) the 
collision, but the known family members are always tightly clustered dynamically. Figure 1 shows the extent of proper 
element space covered by a collision with $\Delta v_{min}$=150 \mpers; this collisional cloud contains all the known family 
members (allowing for resonance diffusion).

\subsection{Non-resonant KBOs}

Keeping in mind the large spread in osculating elements of KBOs that could belong to the \elso family, 131 
high-inclination KBOs observed over at least two oppositions were chosen for further study. These objects were 
integrated using the n-body code SyMBA \citep{1994Icar..108...18L} using the integrator \verb+swift_rmvs3+ based on the 
mapping by \citet{1991AJ....102.1528W}. The integration proceeded backwards in time with 40-day timesteps from epoch JD 
2451545.0 and included the 4 outer planets and the KBOs as test particles with initial conditions given by JPL 
HORIZONS\footnote{\texttt{http://ssd.jpl.nasa.gov/horizons.cgi}}.
Proper elements were taken as the 50 MYr average of the corresponding osculating elements. 

Using the center of mass orbit found above reveals that 2003 UZ117 and 
\Nfive have small values of 
$\Delta v_{min}$, less than some known fragments. (No other KBOs have 
significantly smaller $\Delta v_{min}$ than known family members.) KBO 
2003 UZ117 has unpublished colors obtained by Tegler et 
al.\footnote{\texttt{http://www.physics.nau.edu/$\tilde{}$ 
tegler/research/survey.htm}} that show it has a clearly neutral color 
gradient. As shown in \bfamns, all family members have blue/neutral 
visible color gradients (see Table 1 and references therein). Although 
future infrared observations are necessary, since this object has a 
strongly consistent color and is dynamically within the core of other 
known fragments, we will consider it a member of the \elso family. No 
color or spectral information is available for 2005 CB79, but it has a 
very 
low $\Delta v_{min}$ and is an excellent candidate family member.

We now seek a meaningful self-consistent value of $\Delta v_{min}$ for other KBOs which may or may not be other family 
members. Allowing the center of mass orbital angles to vary can significantly change the shape of the collisional cloud 
and the values of $\Delta v$ as illustrated in Figure \ref{dvexplain}. We could use the center of mass orbit found 
above (thin lines in Figure \ref{dvexplain}), but the small number of family members and their tight clustering does 
not provide a unique constraint for the orbital angles. On the other hand, the orbital angles that minimize $\Delta v$ 
for each individual KBO may result in a collisional cloud that is inconsistent with the distribution of known family 
members (dotted line in Figure \ref{dvexplain}). As a compromise, we define $\Delta v_{min}$ for candidate KBOs as the 
minimum $\Delta v$ found by varying the orbital angles under the constraint that all known family members must lie 
within the resulting collisional cloud (thick solid curve in Figure \ref{dvexplain}). In other words, the angles are 
allowed to vary so long as the candidate KBO has larger $\Delta v_{min}$ than all the known family members. This compromise 
allows for flexibility in estimating the center of mass orbital angles that are compatible with the known family 
members.

The results of this analysis are given in Table 1. For those KBOs known to 
be family members, accounting for errors in the orbital elements (as 
listed on the AstDys 
website\footnote{\texttt{http://hamilton.dm.unipi.it/cgi-bin/astdys/astibo}}) 
caused variations in $\Delta v_{min}$ of less than 5-10\%.

\clearpage

\begin{deluxetable}{lrllllllll}
\tabletypesize{\scriptsize}
\tablewidth{0pt}
\rotate
\tablecaption{KBOs Near the \elso Family}

\tablehead{
\colhead{Name} & \colhead{$\Delta v_{min}$ (\mpers)} & \colhead{$a_P$ (AU)} & \colhead{$e_P$} & \colhead{$i_P (^{\circ})$}
 & \colhead{$T_P$} & \colhead{H\tablenotemark{a}} & \colhead{Visible Gradient\tablenotemark{b}} 
 & \colhead{Comments on Infrared Spectra} & \colhead{References}  }

\startdata

 1996 TO66 &   24.2 &  43.32 &   0.12 &  28.02 & 2.83 &  4.50 &
$     2.38 \pm 2.04$ &                    Strong Water Ice &    \bfamt,\nollpostert \\
2003 UZ117 &   66.8 &  44.26 &   0.13 &  27.88 & 2.84 &  5.20 &
$     0.00 \pm 1.96$ &                 (Strong Water Ice)? &             \teglert \\
\Nfive     & 96.7 & 43.27 & 0.13 & 27.17 & 2.84 &  5.0 &
$ NA$                     &                                 &                           \\
2002 TX300 &  107.5 &  43.29 &   0.13 &  26.98 & 2.84 &  3.09 &
$     0.00 \pm 0.67$ &                    Strong Water Ice &               \bfamt \\
 2005 RR43 &  111.2 &  43.27 &   0.13 &  27.07 & 2.84 &  4.00 &
$                NA$ &                    Strong Water Ice &               \bfamt \\
 2003 OP32 &  123.3 &  43.24 &   0.10 &  27.05 & 2.85 &  4.10 &
$    -1.09 \pm 2.20$ &                    Strong Water Ice &               \bfamt \\
  2005 FY9 &  141.2 &  45.56 &   0.16 &  27.63 & 2.84 & -0.23 &
$                NA$ &                  Methane Ice &             \fyninet \\
 2002 GH32 &  141.9 &  42.04 &   0.09 &  27.59 & 2.83 &  5.50 &
$   35.25 \pm 10.21$ &                                     &              \mbosst \\
1998 HL151 &  142.5 &  40.80 &   0.09 &  27.82 & 2.82 &  8.10 &
$     9.83 \pm 21.2$ &                                     &              \mbosst \\
2003 SQ317 &  148.0 &  42.67 &   0.09 &  28.16 & 2.83 &  6.30 &
$                NA$ &                                     &                      \\
 1995 SM55 &  149.7 &  41.84 &   0.10 &  26.85 & 2.84 &  4.80 &
$     1.79 \pm 2.60$ &                    Strong Water Ice &       \bfamt,\nollpostert \\
  1999 OK4 &  161.5 &  43.30 &   0.15 &  28.58 & 2.81 &  7.60 &
$                NA$ &                                     &                      \\
2004 PT107 &  198.3 &  40.60 &   0.06 &  27.32 & 2.83 &  5.60 &
$                NA$ &                                     &                      \\
\Kfive     & 199.2 & 43.46 & 0.16 & 27.12 & 2.84 &  3.7 &
$ NA$                     &           Weak Water Ice         &     \barkumet            \\
 2003 HA57 &  214.3 &  39.44 &   0.15 &  28.40 & 2.78 &  8.10 &
$                NA$ &                                     &                      \\
 2004 SB60 &  221.0 &  42.08 &   0.10 &  25.59 & 2.86 &  4.40 &
$                NA$ &                                     &                      \\
 2003 TH58 &  229.6 &  39.44 &   0.06 &  29.50 & 2.78 &  7.60 &
$                NA$ &                                     &                      \\
 1998 WT31 &  233.3 &  46.04 &   0.19 &  27.91 & 2.83 &  7.05 &
$     5.57 \pm 5.61$ &                                     &              \mbosst \\
2002 AW197 &  265.0 &  47.28 &   0.12 &  26.00 & 2.90 &  3.27 &
$    22.45 \pm 1.62$ &                        No Water Ice &               \bfamt \\
 1996 RQ20 &  269.9 &  43.89 &   0.10 &  31.74 & 2.76 &  6.95 &
$    19.81 \pm 6.31$ &     IR Colors Inconsistent with Family &        \mbosst,\nollpostert \\
  1999 OY3 &  292.8 &  43.92 &   0.17 &  25.80 & 2.86 &  6.76 &
$    -2.62 \pm 3.39$ & Vis and IR Colors of Strong Water Ice & \mbosst,\nollpostert \\
  1999 OH4 &  305.1 &  40.52 &   0.04 &  26.71 & 2.84 &  8.30 &
$                NA$ &    IR Colors Inconsistent with Family &  \nollpostert      \\
  1997 RX9 &  306.1 &  41.62 &   0.05 &  29.31 & 2.80 &  8.30 &
$                NA$ &                                     &                      \\
2001 QC298 &  310.2 &  46.32 &   0.13 &  31.59 & 2.78 &  6.09 &
$                NA$ &    IR Colors Inconsistent with Family &  \nollpostert       \\
 2003 EL61\tablenotemark{c} &  323.5 &  43.10 &   0.19 &  26.85 & 2.83 &  0.27 &
$    -0.18 \pm 0.67$ &                    Strong Water Ice &               \bfamt \\
2000 CG105 &  330.6 &  46.38 &   0.04 &  29.43 & 2.84 &  6.50 &
$    2.58 \pm 17.72$ &    IR Colors Inconsistent with Family &  \mbosst,\nollpostert \\
 2003 HX56 &  363.2 &  47.32 &   0.21 &  30.00 & 2.79 &  7.10 &
$                NA$ &                                     &                      \\
1999 CD158 &  364.0 &  43.71 &   0.15 &  23.83 & 2.90 &  5.05 &
$    16.36 \pm 3.41$ &  IR Colors Inconsistent with Family &     \mbosst,\nollpostert \\

\enddata

\tablerefs{
(\fyninet) \citet{2007AJ....133..284B} \quad
(\barkumet) Barkume et al. in press \quad
(\bfamt) \citet{\bfamnat} \quad
(\mbosst) MBOSS Database \texttt{http://www.sc.eso.org/ $\tilde{}$ohainaut/MBOSS/} and references therein 
\quad
(\nollpostert) \citet{2005DPS....37.5611N} \quad
(\teglert) Tegler et al. website \texttt{http://www.physics.nau.edu/ $\tilde{}$tegler/research/survey.htm} \quad
}

\tablecomments{As explained in the text, $\Delta v_{min}$ is the minimum ejection 
velocity required to reach the orbit of the listed KBO from the modeled \elso family-forming collision. The actual ejection velocities could be different, but the relative order should be roughly correct. Known family 
members have visible color gradients near zero and strong water ice spectra. Other objects listed could be family members 
or interlopers.}

\tablenotetext{a}{Absolute Magnitude}
\tablenotetext{b}{As defined in References 1 and \mbosst. Objects without published colors list "NA".}
\tablenotetext{c}{This refers to the current proper elements, without allowing diffusion.}

\end{deluxetable}

\clearpage

The blue/neutral visible color gradients of 1998 HL151 and 1998 WT31 are 
similar to known family members (see Tables 1 and 2 and references 
therein). Blue colors are suggestive, but do not necessarily imply the 
strong water ice spectrum that characterizes this family (\bfamns). 
Without observational evidence of a water-ice spectrum, we cannot confirm 
whether these objects are \elso family members or merely interlopers, 
which appears to be the case for 2002 GH32 and others which have red 
visible color gradients. Table(s) 1 (and 2) then serves as a guide for 
future observations.

\subsection{Resonant KBOs}

As a consequence of the wide dispersion of fragments from a collision in the Kuiper belt, many objects can be injected 
into various mean-motion resonances with Neptune. While KBOs in low-order resonances (e.g. 3:2) can be stable for the 
age of the solar system, objects in high-order resonances (found throughout the region of the \elso family) will 
experience chaotic diffusion, as discussed above. Over timescales of tens of millions to billions of years, the 
proper eccentricity and inclination of resonant KBOs are not conserved. These objects can not be directly connected to 
the family based on current proper elements because their proper elements have changed since the formation of the 
family. How then can we identify such fragments? In the case of \elsons, it is the consistent strong water ice 
spectrum, as well as several indications of a past giant impact, that connect it to the non-resonant objects. 
Similarly, in the asteroid belt, the Eos family intersects the 4:9 Jovian resonance and objects in the resonance have 
diffused in eccentricity and inclination \citep{1995Icar..118..132M}. Spectroscopic studies of a few asteroids in the 
resonance showed them to be uniquely identifiable and consistent with the rest of the Eos family, confirming that these 
fugitives are collisionally linked \citep{2000Icar..145....4Z}.

For resonant KBOs that have not yet diffused to scattered or low-perihelion orbits, the Tisserand parameter with 
respect to Neptune, $T$, can be used as a reasonable dynamical criterion for family membership. Through forward 
modeling, we have verified that the velocity dispersion ($\Delta v \lesssim 300$ \mpers) due to the collision and the 
forced variation of osculating elements in time together cause maximal variations in $T$ of about $\sim 0.1$ from $\sim 2.85$. Only about 16\% of multi-opposition KBOs have a Tisserand parameter between 2.74 and 2.96, and these were included in 
our integrations of KBOs. 

To identify candidate fragments that could have diffused in resonances, we 
allowed the proper eccentricity and inclination of each KBO to vary, 
conserving the proper Tisserand parameter, while the semi-major axis of 
the KBO was fixed to $a_P$. The minimal velocity distance, found using the 
method described above, is called $\delta v_{min}$ to distinguish it from 
the velocity computed with the current proper elements and is listed in 
Table 2. Of course, this value of $\delta v_{min}$ will always be less 
than the corresponding $\Delta v_{min}$ computed with unadjustable proper 
elements and will increase the number of interlopers. Even so, it can give 
an indication of objects that had low ejection velocities and subsequently 
diffused in a resonance. Table 2 lists the resonances present in our 
integration of these objects. Even for multi-opposition KBOs, current 
membership in any of the many weak resonances in this region can be easily 
obscured within the errors in the determination of orbital elements. To be 
conservative, we calculate $\delta v_{min}$ for all KBOs in our 
integration. A lack of resonance identification in Table 2 is not meant to 
imply that these objects have not actually been affected by proper element 
diffusion.

\clearpage

\begin{deluxetable}{lrlllllllll}
\tabletypesize{\scriptsize}
\tablewidth{0pt}
\rotate
\tablecaption{Diffused KBOs Near the \elso Family}

\tablehead{
\colhead{Name} & \colhead{$\delta v_{min}$ (\mpers)} & \colhead{$a_P$ (AU)} & \colhead{$e_{min}$} & \colhead{$i_{min} (^{\circ})$}
 & \colhead{$T_P$} & \colhead{H\tablenotemark{a}} & \colhead{Visible Gradient\tablenotemark{b}} 
 & \colhead{Comments on Infrared Spectra} & \colhead{Resonance} & \colhead{References}  }

\startdata

 1996 TO66 &   15.0 &  43.32 &   0.11 &  28.09 & 2.83 &  4.50 &
$     2.38 \pm 2.04$ &                    Strong Water Ice &       19:11 &      \bfamt,\nollpostert \\
2003 SQ317 &   31.4 &  42.67 &   0.11 &  27.92 & 2.83 &  6.30 &
$                NA$ &                                     &             &                      \\
\Kfive     & 39.0 & 43.27 & 0.12 & 27.77 & 2.84 &  5.0 &
$ NA$                     &          Weak Water Ice         &            &        \barkumet     \\
 2005 RR43 &   58.0 &  43.27 &   0.11 &  27.38 & 2.84 &  4.00 &
$                NA$ &                    Strong Water Ice &             &               \bfamt \\
2003 UZ117 &   60.8 &  44.26 &   0.12 &  28.01 & 2.84 &  5.20 &
$     0.00 \pm 1.96$ &                 (Strong Water Ice)? &             &             \teglert \\
\Nfive     & 66.5 & 43.27 & 0.11 & 27.40 & 2.84 &  5.0 &
$ NA$                     &                                 &            &               \\
2002 TX300 &   68.4 &  43.29 &   0.11 &  27.23 & 2.84 &  3.09 &
$     0.00 \pm 0.67$ &                    Strong Water Ice &             &               \bfamt \\
  1999 OK4 &   72.5 &  43.30 &   0.12 &  29.16 & 2.81 &  7.60 &
$                NA$ &                                     &             &                      \\
 2002 GH32 &   79.3 &  42.04 &   0.10 &  27.50 & 2.83 &  5.50 &
$   35.25 \pm 10.21$ &                                     &             &              \mbosst \\
  1997 RX9 &   86.8 &  41.62 &   0.13 &  28.46 & 2.80 &  8.30 &
$                NA$ &                                     &             &                      \\
 2003 OP32 &   91.4 &  43.24 &   0.11 &  26.90 & 2.85 &  4.10 &
$    -1.09 \pm 2.20$ &                    Strong Water Ice &             &               \bfamt \\
  1999 OY3 &   96.6 &  43.92 &   0.10 &  27.00 & 2.86 &  6.76 &
$    -2.62 \pm 3.39$ & IR Colors of Strong Water Ice & \tablenotemark{c} & \mbosst,\nollpostert \\
  2005 FY9 &  118.0 &  45.56 &   0.15 &  27.87 & 2.84 & -0.23 &
$                NA$ &                  Methane Ice &             &             \fyninet \\
 1995 SM55 &  123.3 &  41.84 &   0.09 &  26.98 & 2.84 &  4.80 &
$     1.79 \pm 2.60$ &                    Strong Water Ice &             &    \bfamt,\nollpostert \\
1998 HL151 &  136.4 &  40.80 &   0.11 &  27.55 & 2.82 &  8.10 &
$     9.83 \pm 21.2$ &                                     &             &              \mbosst \\
 1998 WT31 &  139.8 &  46.04 &   0.16 &  28.57 & 2.83 &  7.05 &
$     5.57 \pm 5.61$ &                                     &             &              \mbosst \\
2000 CG105 &  149.0 &  46.38 &   0.16 &  28.04 & 2.84 &  6.50 &
$    2.58 \pm 17.72$ &  IR Colors Inconsistent with Family  &             &    \mbosst,\nollpostert \\
2004 PT107 &  161.9 &  40.60 &   0.09 &  27.08 & 2.83 &  5.60 &
$                NA$ &                                     &             &                      \\
1999 RY215 &  183.0 &  45.28 &   0.11 &  26.37 & 2.88 &  6.13 &
$     4.54 \pm 6.65$ &  IR Colors Inconsistent with Family &             &      \mbosst,\nollpostert \\
2001 FU172 &  200.0 &  39.44 &   0.08 &  28.62 & 2.80 &  8.30 &
$                NA$ &                                     &         3:2 &                      \\
  1999 OH4 &  200.5 &  40.52 &   0.08 &  26.45 & 2.84 &  8.30 &
$                NA$ &  IR Colors Inconsistent with Family  &             &      \nollpostert    \\
 2003 HA57 &  212.3 &  39.44 &   0.12 &  28.80 & 2.78 &  8.10 &
$                NA$ &                                     &         3:2 &                      \\
 2003 TH58 &  214.7 &  39.44 &   0.13 &  28.82 & 2.78 &  7.60 &
$                NA$ &                                     &         3:2 &                      \\
 2004 SB60 &  218.5 &  42.08 &   0.10 &  25.63 & 2.86 &  4.40 &
$                NA$ &                                     &             &                      \\
 2003 QX91 &  222.0 &  43.71 &   0.12 &  31.03 & 2.77 &  8.30 &
$                NA$ &                                     &         7:4 &                      \\
 2000 JG81 &  235.1 &  47.77 &   0.12 &  27.42 & 2.88 &  9.10 &
$                NA$ &                                     &         2:1 &                      \\
 1999 KR16 &  242.9 &  49.00 &   0.22 &  28.34 & 2.84 &  5.70 &
$    44.74 \pm 3.21$ & IR Colors Inconsistent with Family &             &              \mbosst \\
2005 GE187 &  243.5 &  39.44 &   0.07 &  26.50 & 2.84 &  7.10 &
$                NA$ &                                     &         3:2 &                      \\
 1996 TR66 &  248.3 &  47.78 &   0.11 &  26.94 & 2.89 &  7.50 &
$                NA$ &  IR Colors Inconsistent with Family  &             &       \nollpostert   \\

\enddata

\tablerefs{
(\fyninet) \citet{2007AJ....133..284B} \quad
(\barkumet) Barkume et al. in press \quad
(\bfamt) \citet{\bfamnat} \quad
(\mbosst) MBOSS Database \texttt{http://www.sc.eso.org/ $\tilde{}$ohainaut/MBOSS/} and references therein 
\quad
(\nollpostert) \citet{2005DPS....37.5611N} \quad
(\teglert) Tegler et al. website \texttt{http://www.physics.nau.edu/ $\tilde{}$tegler/research/survey.htm} \quad
}

\tablecomments{As explained in the text, $\delta v_{min}$ is the minimum ejection velocity required to reach an orbit with the same proper semi-major axis ($a_P$) and proper Tisserand parameter ($T_P$) of the listed KBOs. Integrations indicate $T_P$ is nearly conserved during eccentricity and inclination diffusion in mean-motion resonances. By construction, \elso is the center of the collision (allowing diffusion). For those objects which are resonant in our integration (identified by libration of the resonance angle in the past 4 MYr), the resonance type is listed. Many more objects may be in 
resonance within the errors of orbit determination (e.g. 1999 OY3), which were not accounted for here.}

\tablenotetext{a}{Absolute Magnitude}
\tablenotetext{b}{As defined in References 1 and \mbosst. Objects without published colors list "NA".}
\tablenotetext{c}{1999 OY3 is probably affected by the 7:4 resonance.}

\end{deluxetable}

\clearpage

As with non-resonant objects, good spectroscopic evidence is required to 
consider these objects part of the \elso family. KBO 1999 OY3 has reported 
near-infrared colors consistent with other family members, which have 
unique colors due to the presence of strong water ice absorptions 
\citep{2005DPS....37.5611N}. Integrations of clones of this KBO that 
include errors in orbital elements show that there is a high probability 
that it is in the 7:4 mean motion resonance. It also has a very low 
resonant $\delta v_{min}$ and, like \elso and 1996 TO66, appears to be a 
family member in a resonance. \Kfive has a very low $\delta v_{min}$, but 
does not have the characteristic spectral features of the family members.

\section{Age of the Family}

Determining the age of the \elso family will allow new insights into the history of the Kuiper belt. It is possible to 
constrain the age because the shape of a collisional cloud will evolve in time by resonance diffusion \citep{1994Natur.370...40M}. In each 
resonance, the changing eccentricity distribution of KBOs can be computed by numerical integrations. (Throughout this 
section, all orbital elements refer to proper orbital elements.) As resonance diffusion is chaotic, ages cannot be 
calculated by back-integration of known particles. Instead, an ensemble of particles (with an assumed initial 
distribution of eccentricities) is integrated forward for the age of the solar system. Comparing the eccentricities of 
remaining particles to the eccentricities of known family members results in an age estimate.

\subsection{Diffusion Time of \elso}

One estimate of the age of the family is the time needed for \elso to diffuse from its original eccentricity to the 
current value. Matching the distribution of known family members above has yielded an estimate of the initial 
eccentricity and inclination of \elso before diffusion, assuming that its displacement from the center of the family is 
small (which is expected from conservation of momentum). From an ensemble of randomly placed test particles, 78 
particles in the 12:7 resonance with low values of $\Delta v$ from the collision center (near $e_{P,orig}=.118$) were 
chosen for long-term integration. Again using SyMBA with a 40-day timestep, these particles were integrated for the age 
of the solar system. The initial configurations of the planets were also randomly chosen (by randomly choosing the 
starting epoch in the past 100 MYr). As expected from the chaotic nature of resonance diffusion, the results do not 
significantly depend on the orientation of the planets or even the initial location of \elsons. Initial eccentricities 
ranged from 0.09 to 0.14, but other than a slight preference for higher eccentricity particles to escape sooner, the 
calculated diffusion times were similar. 

The current proper perihelion of \elso is 35 AU, the approximate limit for stability against close encounters with 
Neptune. In our simulations, particles are usually removed shortly after attaining the current eccentricity of EL61 
$e_{P,now}=0.186$, though they occasionally diffuse back down to lower eccentricities. In Figure \ref{ageplot}, we have plotted the 
fraction of particles with proper eccentricities below $e_{P,now}$ as a function of time. Nearly 90\% of the particles 
have not diffused the full distance a billion years into the integrations. After 3.5 billion years of evolution, 
roughly half of the particles have reached the current eccentricity of \elsons. We conclude that with 90\% confidence, 
the \elso family is older than 1 GYr, with an age estimate of 3.5 $\pm$ 2 GYr (1-sigma). Importantly, the age is 
completely consistent with formation at the beginning of the solar system: the family is ancient and likely primordial.

\subsection{Diffusion Time of Other Resonant Family Members}

As the progenitor of the collision, the resonance diffusion of \elso is a special case because it is a unique object. For other family members, a similar analysis would require assumptions about the number of similar objects captured in each resonance. To avoid unnecessary assumptions, we propose a simple 
method for estimating the age of the family by using the eccentricity distribution of family members currently in the 
resonance instead of focusing on any single particle. The initial eccentricity distribution in a resonance can be inferred from the eccentricity distribution of 
nearby non-resonant particles whose (proper) eccentricities are essentially constant for the age of the solar system. Integrating 
an ensemble of particles with the same starting eccentricities will produce an evolving eccentricity distribution. As 
an example, consider the eccentricity evolution for the 12:7 resonants shown in Figure \ref{el61ageeccdistfig}; only 
the eccentricities of objects still in the resonance are shown. After debiasing the eccentricity distribution of known 
family members for detection biases, the current eccentricity distribution can be statistically compared with the 
integrations. As it uses only the distribution of remaining eccentricities, this method does not require any 
assumptions about the efficiency of initial emplacement in or removal from the resonance in question.

The characteristic diffusion timescale of these resonances strongly depends on the order or strength of the resonance. 
While \elso is in the fifth-order 12:7 resonance, 1996 TO66 appears to be in the weaker eighth-order 19:11 resonance. 
Over the age of the solar system, the proper eccentricities of 19:11 resonants (near the current location of TO66) only 
change by 0.01-0.02; this is less than the accuracy with which one could infer the unknown initial location of these 
objects. The eccentricity distribution of objects in such weak resonances are essentially constant in time and therefore cannot provide a meaningful age constraint.

In contrast, 1999 OY3 is in the strong third-order 7:4 resonance. In addition to increased diffusion times, 7:4 
resonants often participate in the Kozai resonance which allows for exchange of angular momentum between eccentricity and inclination \citep{2005P&SS...53.1175L}. Particles near the assumed starting position of OY3 
unpredictably enter and leave the Kozai resonance, potentially causing huge swings in eccentricity (as these are all high 
inclination particles) on very short (MYr) timescales. Practically any current eccentricity could have possibly been generated by the Kozai resonance in less than 10 MYr. Therefore, this resonance also appears to be a fruitless source 
of useful age constraints from eccentricity diffusion. It is important to note here that the assumption of constant 
$T_P$ is often violated by these particles; the Kozai resonance carries them to low perihelia where they interact with 
Neptune (which is itself interacting with the other giant planets) often causing a change in $T_P$. Therefore, KBOs 
with low perihelia in the past or present are susceptible to inaccurate estimations of the initial ejection velocity 
($\delta v_{min}$). Although these particles can be significantly perturbed, in an integration of 20 particles in the 7:4 
resonance starting near the estimated initial conditions of OY3, half are still in the resonance after 4.5 GYr.

\subsection{Future Age Estimates}

In the future, more family members will be identified in resonances that will, like the 12:7 resonance, have diffusion timescales that are neither too slow nor too fast. The addition of several new family members will allow for further refinement of these age estimates. 

It is possible to evaluate the number of resonant particles needed to make a significant improvement upon our age estimate of 3.5 $\pm$ 2 GYr. From the distributions of remaining 12:7 resonants shown in Figure \ref{el61ageeccdistfig}, we randomly drew 5, 10, 50, and 100 eccentricities with replacement at half-billion year intervals from particles that were still in the resonance. Every pair of distributions was inter-compared using the Kuiper variant of the K-S test which returns a significance near 1 if the distributions are distinguishable statistically. This process (of random selection and cross-comparison) was repeated 100 times and the results averaged. Figure \ref{eccdistcorfig} shows some of the results of this analysis. For example, if the actual age of the family is 3.5 GYr (Figure \ref{eccdistcorfig}, upper right) then with $\sim$50 particles (triangles), an age of 2.5 GYr or younger can be ruled out since these eccentricity distributions are different with greater than 90\% significance. With $\sim$100 resonant particles, the accuracy in the age determination can generally be brought down to 0.5 GYr. 

Unfortunately, the chaotic nature of eccentricity diffusion makes it difficult to determine a precise age without a large number of objects. However, it should be noted that each resonance gives an essentially independent measure of the age so that a total of 50-100 known family members in the appropriate (e.g. not too fast or too slow) resonances should be sufficient to get a relatively precise age estimate. In the near future, as the high-inclination region of the Kuiper belt is probed much more deeply, many new family members should be readily discovered. 

\section{Discussion}\label{disc}

Identifying more \elso family members is very useful for learning more about this family and its relationship to the 
formation and evolution of the outer solar system. Based on dynamical and observational evidence, we add 2003 UZ117 and 
1999 OY3 to the list of \elso family members, although the former should be observed in the infrared for confirmation 
of a strong water ice signature.

Due to the highly dispersive nature of large Kuiper belt collisions, some simplifications were made to identify 
potential family members. The computed values of $\Delta v_{min}$ (Table 1) and especially $\delta v_{min}$ (Table 2) could be 
significantly different from the true ejection velocities. Even so, we find it highly significant that all of the known 
fragments of the \elso family can be explained by a velocity dispersion of 150 \mpersws from a single collision location, 
and allowing the objects in resonances to diffuse in eccentricity. In addition, all known KBOs near the proposed 
collision have strong water ice signatures, including the strongest such absorption features known in the Kuiper belt 
(except possibly 2003 UZ117, whose spectrum is unknown) (\bfamns). Combining dynamical and spectroscopic evidence, the 
\elso family currently includes, in order of decreasing absolute magnitude: (136108) 2003 EL61, (55636) 2002 TX300, 
(145453) 2005 RR43, (120178) 2003 OP32, (19308) 1996 TO66, (24835) 1995 SM55, 2003 UZ117, and (86047) 1999 OY3.

Many potential family members have no known photometric or spectroscopic observations. Observations of near-infrared 
colors on these objects will help to distinguish family members from interlopers. Discovery of additional family 
members will do much to improve our understanding of this family and the outer solar system. In particular, fragments 
in resonances have the unique ability to constrain the age of the collisional family as eccentricity diffusion, though 
chaotic, is time dependent. Based on the time needed for \elso to diffuse to its current location, the family-forming 
collision occurred at least a billion years ago. Indeed, the probability of such a collision is only reasonable in the 
primordial Kuiper belt when the number densities of large KBOs was much higher. However, the collision should 
have occurred after any significant dynamical stirring as the orbital distribution of the family remains tight and seemingly 
unperturbed. 

There appears to be no dynamical evidence that is not consistent with the 
formation of the \elso family by an ancient collision. It is therefore 
interesting that all family members appear to be bright and pristine with 
strong crystalline water ice spectra. (B07) These surfaces seem to be 
exceptions to the premise that all static surfaces in the outer solar 
system darken and redden in time (i.e. \citet{1996AJ....112.2310L}). This 
is not due to their location in the outer solar system, as there are KBOs 
dynamically nearby with red spectrally-featureless surfaces (see Tables 1 
and 2). Perhaps the collision was energetic enough to sublimate and lose 
volatiles before they were able to transform into the higher-order 
hydrocarbons that are thought to be the darkening reddening agent dominant 
in the outer solar system. However, this does not really distinguish these 
family members from all KBOs, at least some of which should have 
experienced similarly energetic impacts. Instead, the distinguishing 
characteristic may be that the relatively large proto-\elso was able to 
fully differentiate and the resulting fragments were compositionally much 
purer than other objects, even fragments from non-differentiated 
progenitors. In any case, the unique spectra of family members promise to 
improve our understanding of outer solar system surface processes (Barkume 
et al. 2007, in press).

Understanding the surfaces of KBOs is one of many insights provided by the likely primordial nature of the \elso 
family. Another is the apparent need for higher number densities in the past if the family-forming collision is to be rendered probable. Continuing identification and characterization of family members can uniquely improve our understanding of this collision and its connection to the formation and evolution of the outer solar system.

{\it Acknowledgments:} We would like to thank Kris Barkume, Alessandro Morbidelli, Keith Noll, Emily Schaller, Meg Schwamb, and Jack Wisdom for valuable discussions. We would like to thank the referee (David Nesvorn{\'y})for timely reviews that improved the quality of this paper. DR is grateful for the support of the Moore Foundation.

\bibliographystyle{aj}

\clearpage

\begin{figure}
\includegraphics[angle=90,width=6in]{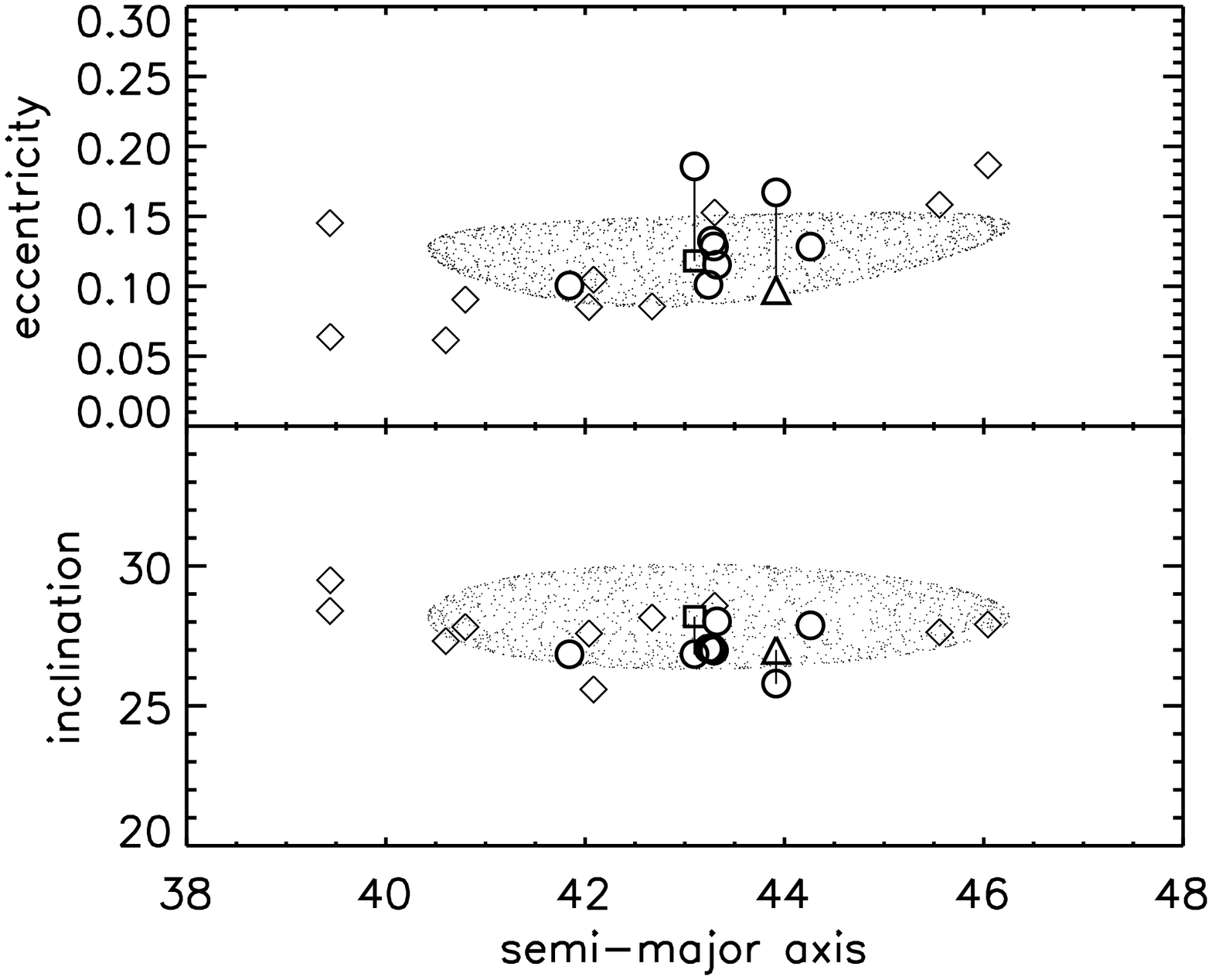}
	    \caption{\label{fig1} Current proper orbtial elements of members of the \elso family (circles) and 
potential family members (diamonds). The cloud of small points illustrate the dispersion in semi-major axis, 
eccentricity, and inclination of objects ejected from the nominal collision location (located at the center of the 
square) with an isotropic $\Delta v$ of 150 \mpers, enough to explain all the currently known members of the \elso 
family. (Note that the orbital angles are chosen to minimize $\Delta v$; the actual ejection velocities may be larger.) The square identifies the calculated location of the collision center which is assumed to be the initial 
location of \elso before resonance diffusion (marked by vertical lines). The two rightmost circles are the current 
proper elements of 1999 OY3 and 2003 UZ117, new family members identified in this work.  KBO 1999 OY3 (which has 
visible and infrared colors consistent with family members) is also allowed to diffuse to the location marked by the 
triangle (see Table 2). The proper elements of other KBOs with $\Delta v <$ 250 \mpers (listed in Table 1) are shown as 
diamonds.  
} 
\end{figure}

\begin{figure}
\includegraphics[angle=90,width=6in]{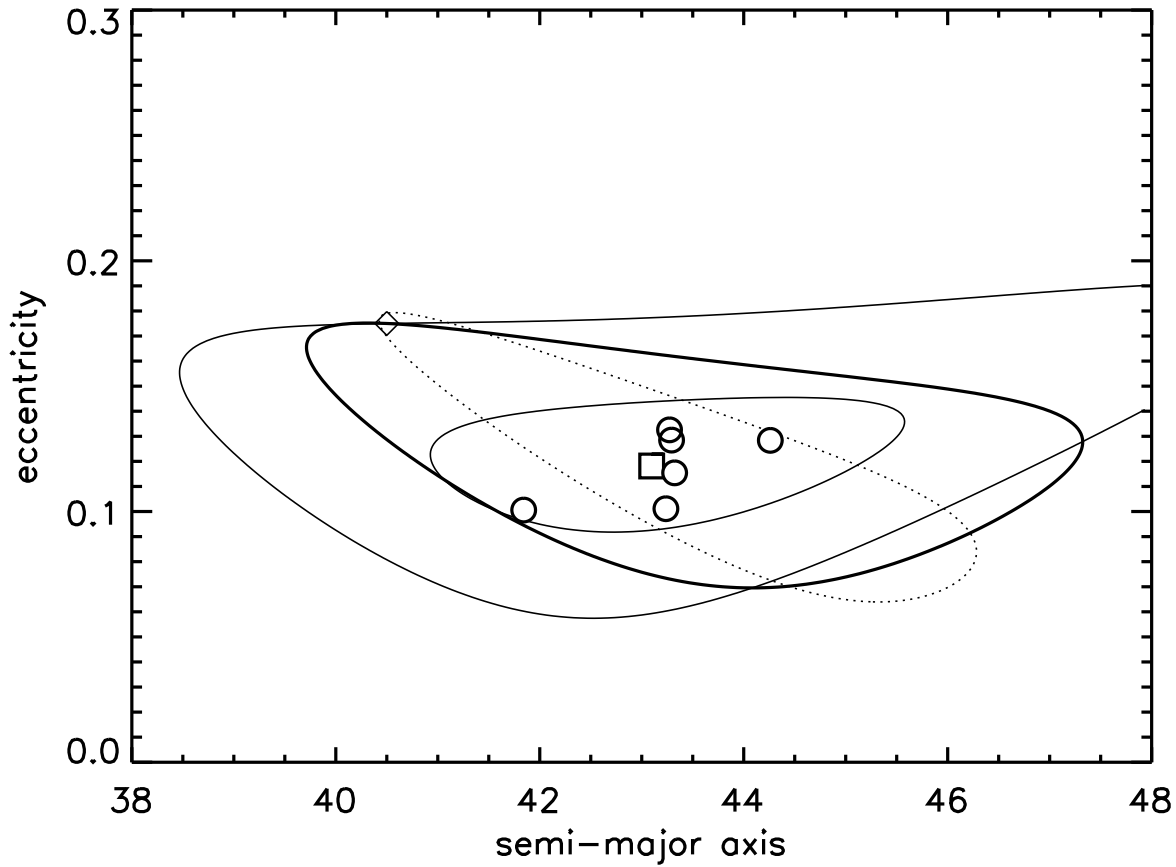}
	    \caption{\label{dvexplain} Illustration of determining $\Delta v$ for candidate family members. The known (non-diffusing) family members (circles) fall into a relatively small region of proper orbital element space. (In this illustration, inclinations are held constant and collisions are located near the ecliptic.) These can be explained by ejection from the center of mass orbit (square) with $\Delta v$ of 150 \mpers or less, marked by the smallest thin solid curve. A hypothetical KBO (diamond) can be explained with larger $\Delta v$ from the same center of mass orbit, shown by the larger thin curve. Alternatively, by changing the orbital angles, particularly the mean anomaly, the shape of the collisional cloud can also change requiring a much smaller $\Delta v$ (dotted curve). However, the resulting angle may be inconsistent with the distribution of known family members, as in the case above. As a compromise between these two methods, we find the $\Delta v_{min}$ for KBOs by allowing the center of mass orbital angles to vary with the constraint that all known family members must lie within the collisional cloud (thick solid curve).
} 
\end{figure}

\clearpage

\begin{figure}
\includegraphics[angle=90,width=6in]{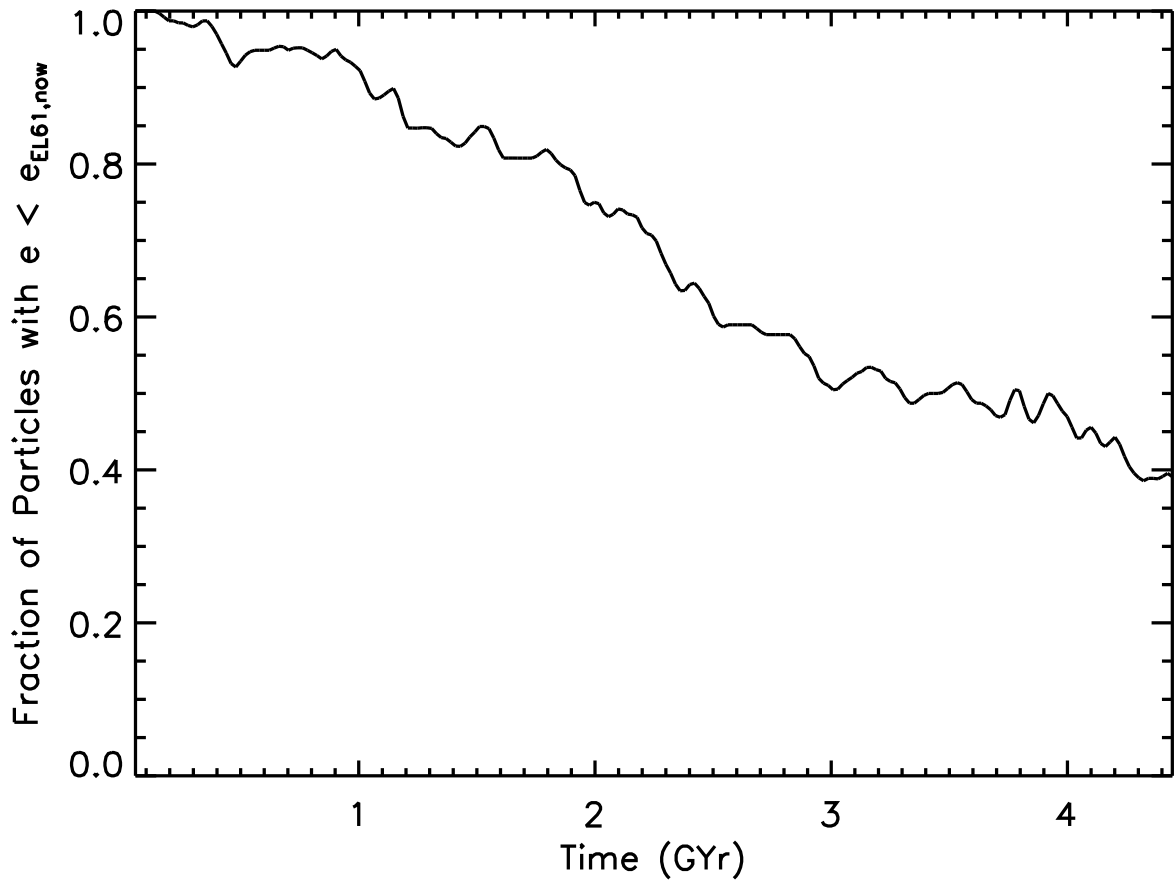}
	\caption{\label{ageplot} Fraction of particles that have eccentricities less than the current eccentricity of 
\elsons $(e_P < 0.186$). Calculated from an integration of 78 12:7 resonants with initial eccentricities near 0.118 
(the expected initial eccentricity of \elsons). The current location of \elso is attained by 10\% of resonant KBOs in 
less than $\sim$1 GYr. After nearly 4 GYr of evolution, half the particles have passed the current location of \elsons. 
We conclude that the \elso family is ancient and probably primordial. }

\end{figure}

\clearpage

\begin{figure}						 
\includegraphics[angle=90,width=6in]{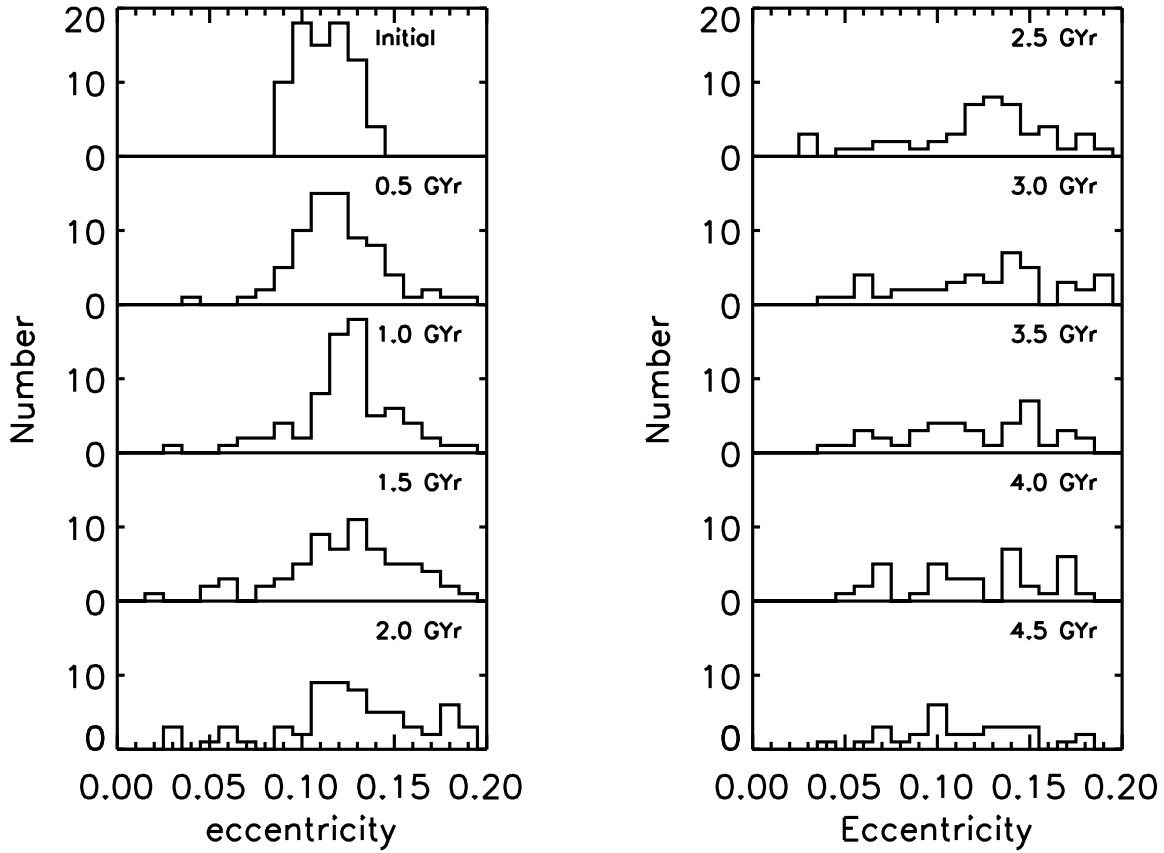}
	\caption{\label{el61ageeccdistfig} Distribution of eccentricities of 12:7 resonants at half-billion year intervals. Only 
particles remaining in the resonance are shown. The bin size is 0.01 in eccentricity. A clear diffusion-like spreading 
is evident in the widening of the initial peak. The eccentricities of KBOs in this resonance will have a 
distribution that can be compared to these distributions in order to estimate an age. 
}

\end{figure}

\clearpage

\begin{figure}
\includegraphics[angle=90,width=6in]{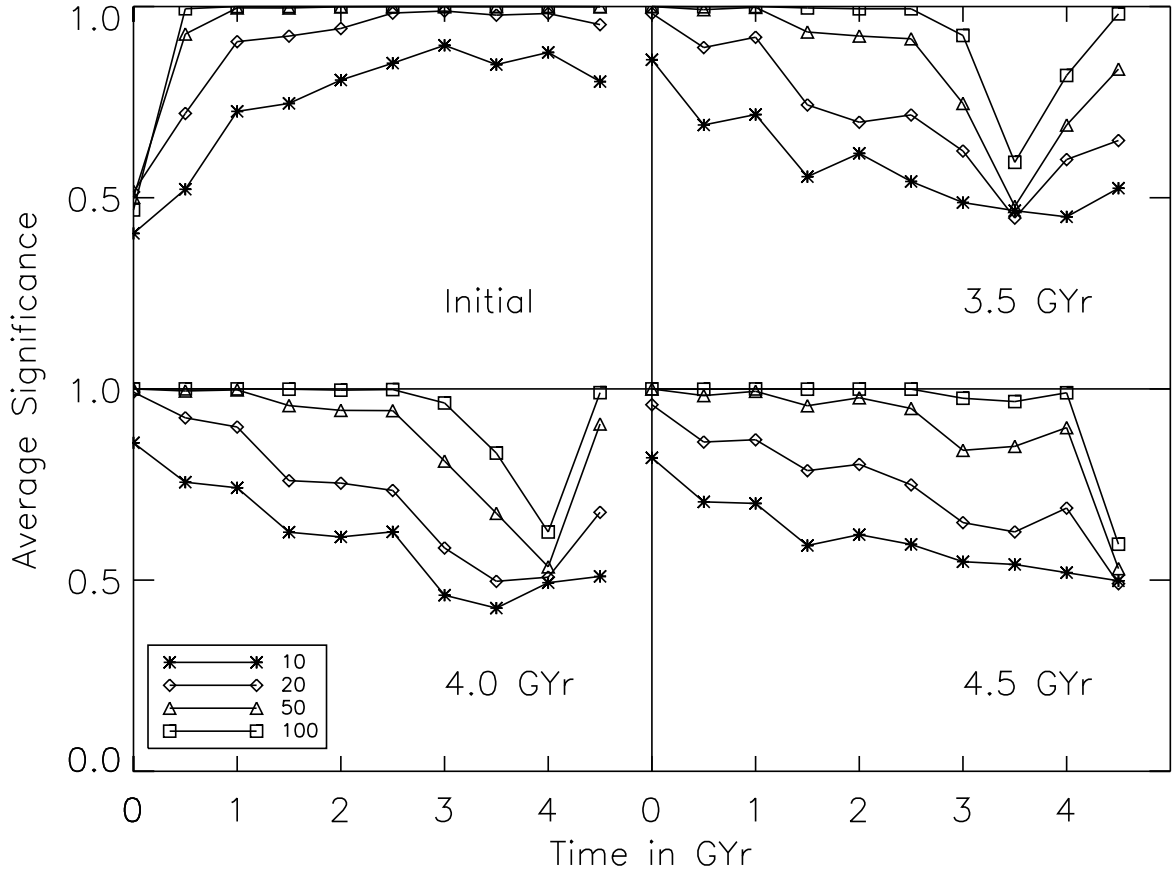}
	\caption{\label{eccdistcorfig} Correlations of eccentricity distributions with each other. As explained in the text, 
a random sample of particles is chosen from two ages and compared using the Kuiper variant of the K-S test. The 
probability that the two distributions are different is represented by the significance. The cross-comparisons with all 
ages are shown for the initial distribution (upper left) and the distributions at 3.5 GYr (upper right), 4.0 GYr (lower left), and 4.5 GYr (lower right). As expected the significance of being drawn from different populations is least when 
the distribution at each age is compared with itself. The different symbols represent the number of particles drawn 
from each distribution as shown in the legend. About 50-100 resonant objects are needed to strongly distinguish ages that differ by only half a billion years.
}

\end{figure}

\clearpage

\end{document}